\documentclass[twocolumn]{aastex63}

\usepackage{url}

\newcommand{\vlsr} {v_\mathrm{lsr}}
\newcommand{\pa} {\mathrm{P.A.}}

\newcommand{\h} {^{\mathrm{h}}}
\newcommand{\m} {^{\mathrm{m}}}
\newcommand{\s} {^{\mathrm{s}}}

\newcommand{\mJybeam}  {\mbox{mJy}~\mbox{beam}^{-1}}

\newcommand{\kkms}  {\mbox{K}~\mbox{km s}^{-1}}

\newcommand{\kms}	{\mbox{km s}^{-1}}

\newcommand{\yr}	{{\rm yr}}
\newcommand{\K}	{{\rm K}}

\newcommand{\au} {\mbox{au}}

\newcommand{\natas} {NatAs}

\shorttitle{Salty Twin Disks}
\shortauthors{Tanaka et al.}

\begin{document}

\title{Salt, Hot Water, and Silicon Compounds Tracing Massive Twin Disks}

\correspondingauthor{Kei E. I. Tanaka}
\email{kei.tanaka@nao.ac.jp}

\author[0000-0002-6907-0926]{Kei E. I. Tanaka}
\affiliation{National Astronomical Observatory of Japan, Mitaka, Tokyo 181-8588, Japan}
\affiliation{Department of Earth and Space Science, Osaka University, Toyonaka, Osaka 560-0043, Japan}

\author[0000-0001-7511-0034]{Yichen Zhang}
\affiliation{Star and Planet Formation Laboratory, RIKEN Cluster for Pioneering Research, Wako, Saitama 351-0198, Japan}

\author[0000-0003-1659-095X]{Tomoya Hirota}
\affiliation{National Astronomical Observatory of Japan, Mitaka, Tokyo 181-8588, Japan}
\affiliation{Department of Astronomical Sciences, SOKENDAI (The Graduate University for Advanced Studies), Mitaka, Tokyo 181-8588, Japan}

\author[0000-0002-3297-4497]{Nami Sakai}
\affiliation{Star and Planet Formation Laboratory, RIKEN Cluster for Pioneering Research, Wako, Saitama 351-0198, Japan}

\author[0000-0002-3789-770X]{Kazuhito Motogi}
\affiliation{Graduate School of Sciences and Technology for Innovation, Yamaguchi University, Yamaguchi 753-8512, Japan}

\author[0000-0001-8105-8113]{Kengo Tomida}
\affiliation{Astronomical Institute, Tohoku University, Sendai 980-8578, Japan}
\affiliation{Department of Earth and Space Science, Osaka University, Toyonaka, Osaka 560-0043, Japan}

\author[0000-0002-3389-9142]{Jonathan C. Tan}
\affiliation{Department of Space, Earth \& Environment, Chalmers University of Technology, SE-412 96 Gothenburg, Sweden}
\affiliation{Department of Astronomy, University of Virginia, Charlottesville, VA 22904-4325, USA}

\author[0000-0001-8596-1756]{Viviana Rosero}
\affiliation{National Radio Astronomy Observatory, 1003 Lopezville Road, Socorro, NM 87801, USA}

\author[0000-0002-9221-2910]{Aya E. Higuchi}
\affiliation{National Astronomical Observatory of Japan, Mitaka, Tokyo 181-8588, Japan}

\author[0000-0002-9661-7958]{Satoshi Ohashi}
\affiliation{Star and Planet Formation Laboratory, RIKEN Cluster for Pioneering Research, Wako, Saitama 351-0198, Japan}

\author[0000-0001-6159-2394]{Mengyao Liu}
\affiliation{Department of Astronomy, University of Virginia, Charlottesville, VA 22904-4325, USA}

\author[0000-0002-6033-5000]{Koichiro Sugiyama}
\affiliation{National Astronomical Research Institute of Thailand, 260 Moo 4, T. Donkaew, A. Maerim, Chiang Mai 50180, Thailand}
\affiliation{National Astronomical Observatory of Japan, Mitaka, Tokyo 181-8588, Japan}

\begin{abstract}
We report results of $0.05\arcsec$-resolution
observations toward the O-type proto-binary system IRAS 16547--4247
with the Atacama Large Millimeter/submillimeter Array.
We present dynamical and chemical structures of the circumbinary disk,
circumstellar disks, outflows, and jets,
illustrated by multi-wavelength continuum and various molecular lines.
In particular, we detect sodium chloride, silicon compounds,
and vibrationally excited water lines as probes of
the individual protostellar disks at a scale of 100 au.
These are complementary to typical hot-core molecules tracing the circumbinary structures on a 1000 au scale.
The H$_2$O line tracing inner disks has an upper-state energy of $E_u/k>3000~\K$,
indicating a high temperature of the disks.
On the other hand, despite the detected transitions of NaCl, SiO, and SiS not necessarily
having high upper-state energies, they are enhanced only in the vicinity of the protostars.
We posit that these molecules are the products of dust destruction,
which only happens in the inner disks. 
This is the second detection of alkali metal halide in protostellar systems
after the case of the disk of Orion Source I,
and also one of few massive protostellar disks associated
with high-energy transition water and silicon compounds.
These new results suggest that these ``hot-disk" lines
may be common in innermost disks around massive protostars,
and have great potential for future research of massive star formation.
We also tentatively find that the twin disks are counter-rotating,
which might give a hint of the origin of the massive proto-binary system IRAS 16547--4247.
\end{abstract}

\keywords{individual objects (IRAS 16547--4247) -- ISM: jets and outflows -- stars: formation -- stars: massive}

\section{Introduction} \label{sec:intro}
Massive stars are the important sources of ultraviolet (UV) radiation, turbulent energy, and heavy elements in galaxies.
Massive close binaries are the progenitors of merging black holes,
which are detected by their gravitational wave emission.
It is of prime importance to understand the formation process of massive stars \citep[e.g.,][]{tan14}.
An essential question is
whether or not massive protostars accrete through disks, as in low-mass star formation.
Recent theoretical/numerical studies support the disk accretion theory \citep[e.g.,][]{ros16,KT17,kui18}.
In particular,
the shielding effect by the disk inside $100{\rm\:au}$
is the key to solving the longstanding radiation pressure problem
\citep[e.g.,][]{wol87}
in the formation of $>40M_\odot$ stars \citep{kui10,KT11}.
Simulations also predict that an accretion disk tends to be gravitationally unstable,
which results in accretion bursts \citep{mat17,mey17,mey18} and the formation of companions \citep{kru09,ros16}.

Thanks to the recent development of interferometers, especially the 
Atacama Large Millimeter/Submillimeter Array (ALMA), more and more
disk/envelope structures around massive protostars with Keplerian-like 
rotation have been reported \citep[see][for recent reviews]{hir18,bel20}.
However, so far, the number of studies reaching the resolution of 
$\la100{\rm\:au}$ remains limited \citep{hir17,gin18,mau19,mot19,joh20}.
The hot and dense nature of the surrounding material of 
massive protostars leads
to the detection of rich molecular lines within $<0.1{\rm\:pc}$,
known as hot cores.
One difficulty in the disk hunting is the lack of knowledge of 
which lines can trace the innermost region and separate the 
disk from the envelope.
Recently, there have been some attempts to identify the disk with both kinematics
and chemical patterns \citep[e.g.,][]{zha19a}.
However, there is no agreed-upon set of such molecular lines.
This work will provide a tip for the disk-tracing line selection 
based on new ALMA observations.

Our target IRAS 16547--4247 (hereafter IRAS 16547) is an O-type protostellar object
with a bolometric luminosity of $\sim10^5\:L_\odot$,
embedded in a $10^3\:M_\odot$ clump within a radius of $0.2{\rm\:pc}$,
at the distance of $2.9{\rm\:kpc}$ \citep{gar03}.
Radio observations showed jets aligned in a northwest--southeast direction,
across a scale of $0.1{\rm\:pc}$ on the plane of the sky \citep{rod05,rod08}.
The presence of jets indicates ongoing accretion in the vicinity of the protostar.
Recently, \citet{zap15,zap19} reported a binary system seen as compact dusty objects with an 
apparent separation of $300{\rm\:au}$, surrounded by a circumbinary disk, using ALMA observations.
Using vibrationally excited CH$_3$OCHO and CS transitions with 
upper-state energies of $E_u/k>500{\rm\:K}$,
\citet{zap19} showed that the circumbinary disk is rotating
with a Keplerian-like profile of an enclosed mass of $25\pm3\:M_\odot$.
However, the dynamics at the several $\times 100~\au$ scale must be
controlled by the individual binary protostars, 
which has not been well studied.

We report new multi-band ALMA observations toward IRAS 16547
with resolutions of $0.05\arcsec$ at 1.3 and 3 mm.
In this Letter, we mainly present the detection of sodium chloride,
silicon compounds, and water lines as probes of the individual circumstellar disks.
We propose that these inner-disk tracers may be common around massive 
protostars at the scale of $\la100{\rm\:au}$,
and valuable in understanding the disk properties
in massive star formation.

\begin{figure*}
\begin{center}
\includegraphics[width=0.8\textwidth]{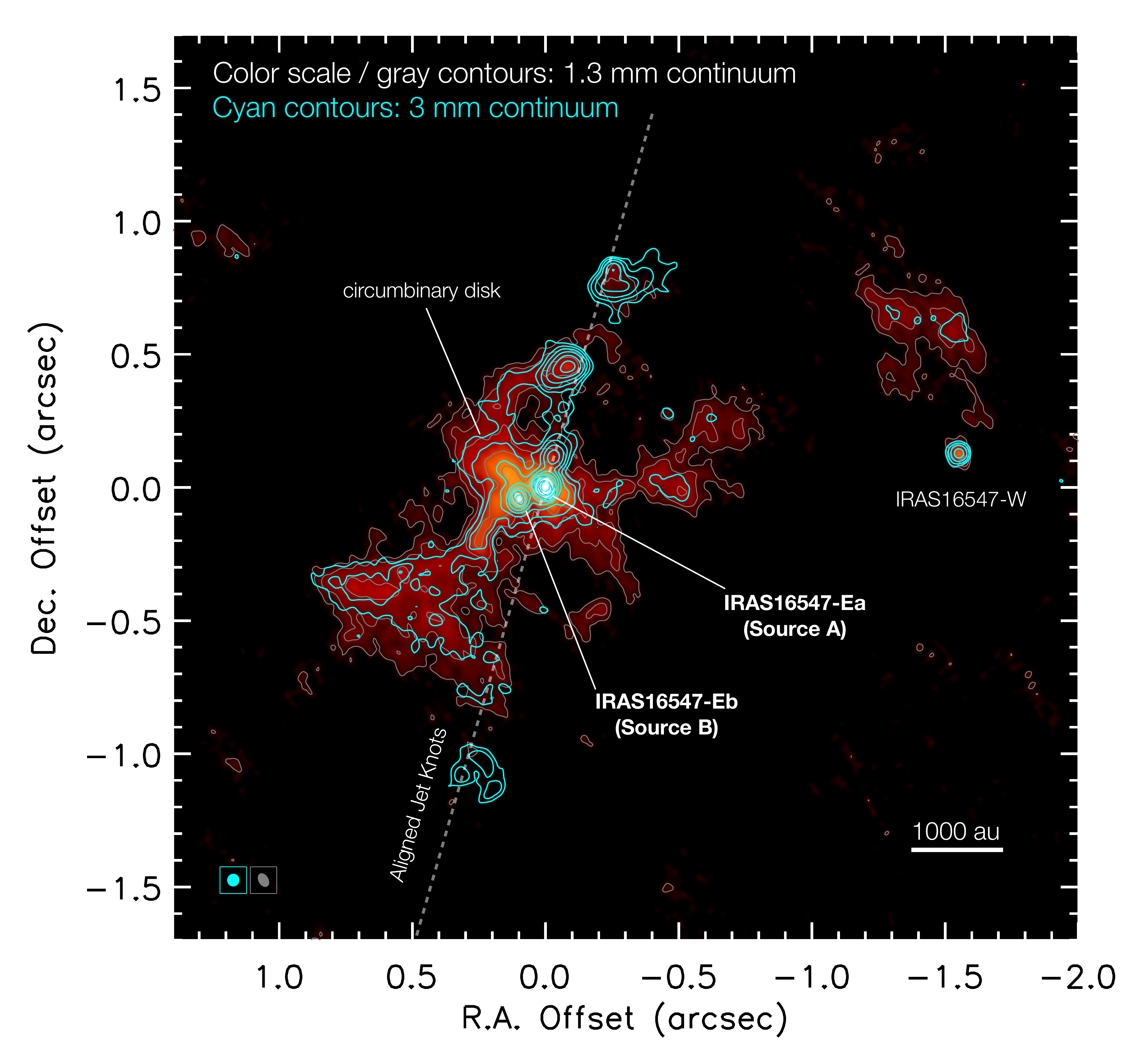}\\
\caption{
ALMA continuum images of IRAS 16547 at
1.3 mm (color scale and gray contours) and 3 mm (cyan contours).
The contour levels are $5\sigma\times2^n$ ($n=0,1,...$), 
with $1\sigma=0.52~\K$ ($0.045~\mJybeam$) for the 1.3 mm continuum,
and $1\sigma=0.54~\K$ ($0.0083~\mJybeam$) for the 3 mm continuum.
The synthesized beams are $0.055\arcsec\times0.038\arcsec$ ($\pa=30.8^\circ$) in the 1.3 mm image
and $0.048\arcsec\times0.046\arcsec$ ($\pa=-16.8^\circ$) in the 3 mm image, respectively (shown at the lower-left corner).
The R.A. and decl. offset is relative to the continuum peak position of source A, i.e.,
$(\alpha,\delta)=\rm(16^h,58^m,17^s.2082,~-42\degr52\arcmin07\arcsec.421)$ (ICRS).
The continuum peak of source B is at
$\rm(16^h,58^m,17^s.2173,~-42\degr52\arcmin07\arcsec.461)$ (ICRS).
}
\label{fig:cont}
\end{center}
\end{figure*}

\begin{figure*}
\begin{center}
\includegraphics[width=\textwidth]{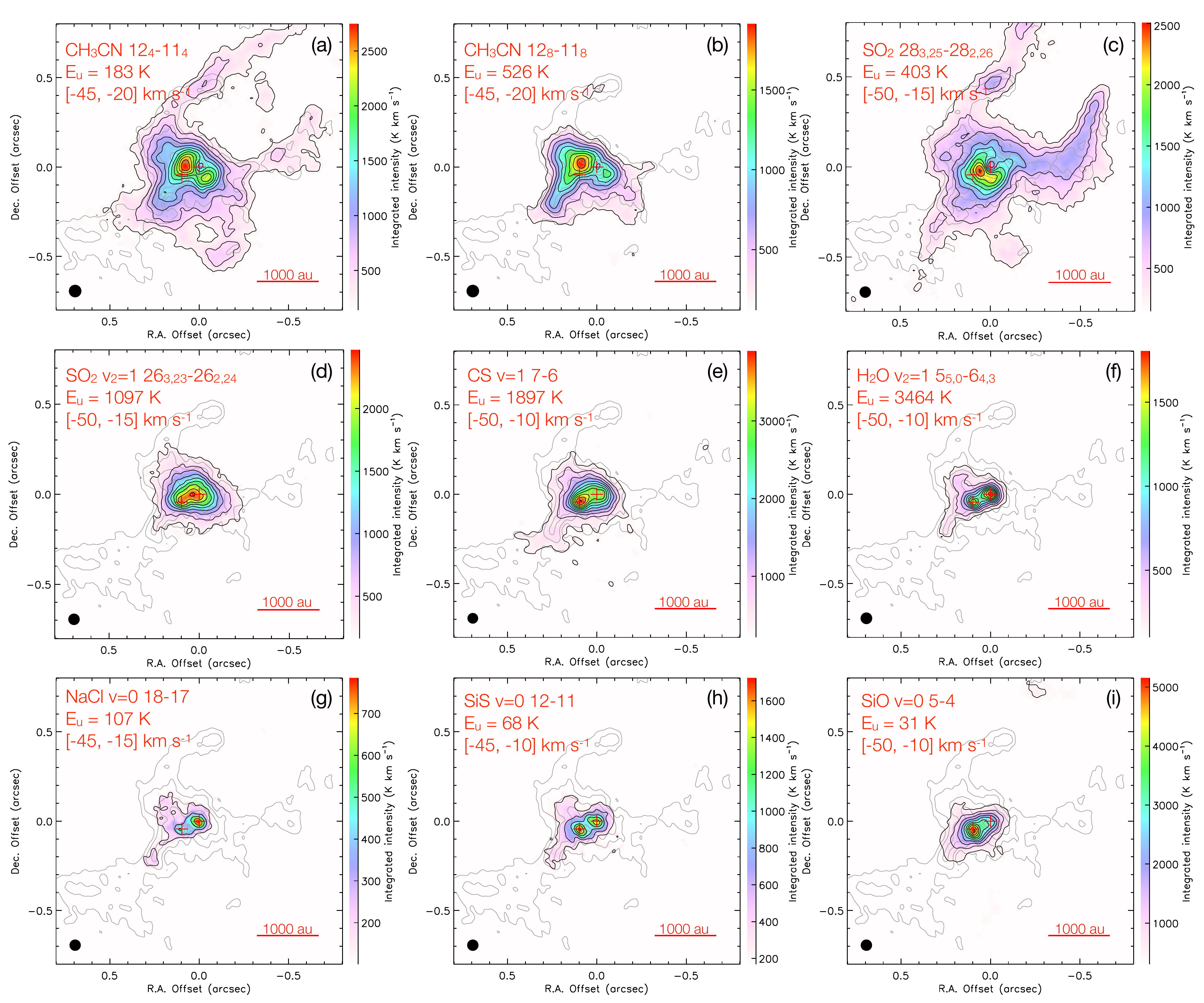}\\
\caption{
Integrated intensity maps of emission lines (color scale and black contours)
overlaid with the 1.3 mm continuum emission (gray contours).
The molecule names, transitions, upper-state energies $E_u$, 
and the integrated $\vlsr$ ranges are labeled in each panel.
The red crosses indicates the continuum peaks of sources A and B.
}
\label{fig:mom0}
\end{center}
\end{figure*}

\section{Observations} \label{sec:obs} 
The 3 mm and 1.3 mm observations were carried out with
ALMA in Band 3 and 6 (ALMA project ID: 2018.1.01656.S).
We also utilize the ALMA archived Band 7 (0.85 mm) data
in project 2016.1.00992.S \citep{zap19}.
We summarize the information of the observations in Appendix Table \ref{tab:obs}.
The data were calibrated using CASA \citep{McMullin07} pipeline v5.6.1.
After pipeline calibration, we performed phase self-calibration 
for all the three bands using the continuum data combining
line-free channels of all the spectral windows, and 
applied the self-calibration solutions to the line data.
Images are made with CASA task tclean
using the Briggs weighting with the robustness parameter of 0.5
for Band 3 and 6 data, and $-0.5$ for Band 7 data.
The resultant synthesized beams of the continuum images
are as high as $0.05\arcsec$ for all wavelengths (Table \ref{tab:obs}).

\section{Results} \label{sec:results}

\subsection{Continuum} \label{sec:continuum}

Figure \ref{fig:cont} shows the 1.3 and 3 mm continuum maps.
The dust emission dominates the 1.3 mm continuum,
highlighting the circumbinary disk and outflow cavities,
while the 3 mm continuum reveals the jet structures.
The structures seen in the 1.3 mm continuum are
very similar to those in 0.85 mm continuum (Appendix Figure \ref{fig:cont2}),
which was first reported by \citet{zap19}.
Three protostars are prominent at all wavelengths,
namely IRAS 16547-Ea and IRAS 16547-Eb 
(hereafter, sources A and B) forming the proto-binary
with an apparent separation of $300{\rm\:au}$,
and a much weaker third source IRAS 16547-W.
Using the 0.85 mm fluxes,
which are less affected by the free-free emission than the 1.3 and 3 mm fluxes,
we evaluate circumstellar disk masses of $0.19\:M_\odot$ and $0.035\:M_\odot$ 
around sources A and B within a radius of $0.05\arcsec$ ($150~\au$)
assuming a dust temperature of $350~\K$ (Appendix \ref{sec:appB}).
The proto-binary is surrounded by a circumbinary disk of $2500{\rm\:au}$,
outflow cavities are seen on the northern and southern sides of the circumbinary disk
\citep[see also][]{zap19}.

The 3 mm continuum newly reveals jet knots from source A aligned in a 
northwest--southeast direction,
which is consistent with
the orientation of the central radio source detected by 
centimeter observations \citep[$\pa=-16\degr$;][]{rod05,rod08}.
The resolution of the ALMA observation is an order of magnitude higher 
than those in the previous radio observations,
which allows us to spatially resolve this central radio source into sources A and B and several jet knots,
and to determine that the jet originates from source A.
The jet orientation is also close to the elongated distribution of water masers \citep{fra09}.
The prominence of the proto-binary and jet knots in 3 mm continuum 
suggests that they are dominated by free-free emissions, and the jet knots 
may also contain significant synchrotron contributions.
We leave the detailed analysis of the multi-band continuum to a forthcoming paper.

\begin{figure*}
\begin{center}
\includegraphics[width=0.95\textwidth]{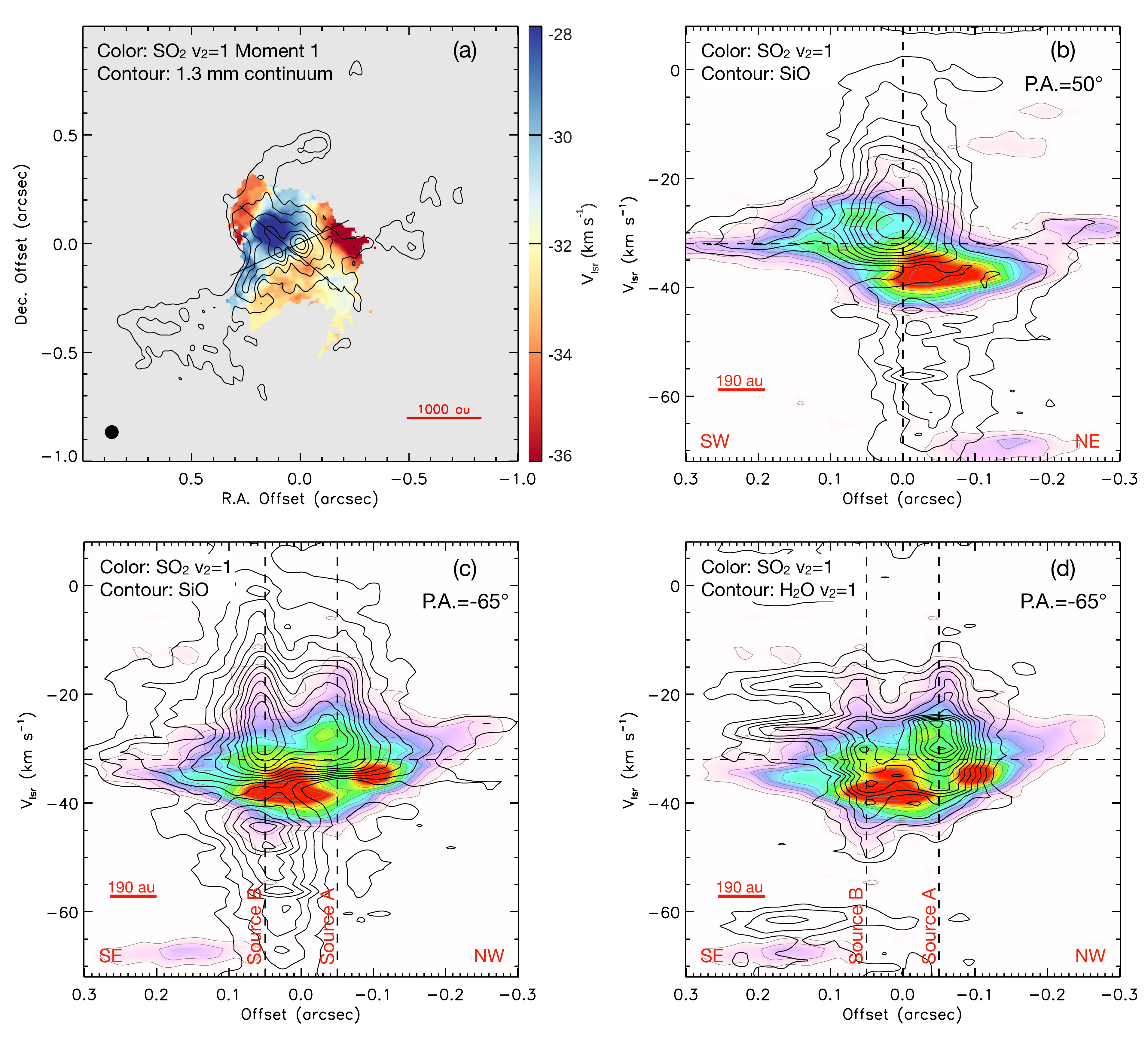}\\
\caption{
{\bf (a)}: Moment 1 map of the SO$_2$ $26_{3,23}-26_{2,24}$ $v_2=1$ line (color scale), 
overlaid with the 1.3 mm continuum maps in contours (same as Figure \ref{fig:cont}(a)).
The two lines show the directions along which the position-velocity (PV) diagrams shown in
panels (b)$-$(d) are made.
{\bf (b)}: The PV diagrams of the SO$_2$ $v_2=1$ line 
(color scale and thin contours) and SiO 5$-$4 line (thick contours),
along a slit passing through the midpoint between of the two sources with $\pa=50^\circ$.
The red bar in the lower-left corner indicates the resolution beam size.
{\bf (c)}: Same as panel (b), but along a slit with $\pa=-65^\circ$.
The locations of the two sources are labeled.
{\bf (d)}: Same as panel (c), but showing the H$_2$O $v_2=1$ line in thick contours.
Note that the PV diagram of H$_2$O is contaminated by other lines around
$\vlsr=-20$ and $-27~\kms$.
In panels (b)$-$(d), the SO$_2$ contours have the first level and intervals of $2.1~\mJybeam$,
the SiO contours have the first level and intervals of $2.2~\mJybeam$,
and the H$_2$O contours have the first level and intervals of $1.4~\mJybeam$.}
\label{fig:outer-disk}
\end{center}
\end{figure*}

\subsection{Lines} \label{sec:lines}

Rich molecular lines are detected in IRAS 16547, 
especially in Bands 6 and 7.
Figure \ref{fig:mom0} shows the integrated intensity maps of 
representative emission lines,
which trace different components in the proto-binary system
from the circumbinary disk to the individual circumstellar disks
(see Appendix \ref{sec:appA} for the summary of the lines presented in this work).
Methyl cyanide CH$_3$CN, which is commonly used as a disk tracer
toward massive protostars \citep[e.g.,][]{joh15,joh20,beu17},
associates with the circumbinary disk and the outflow cavity 
at the 1000 au scale (panels a and b).
We detect the CH$_3$CN $(12_K$--$11_K)$ $K$-ladder 
from $K=0$ to $K=11$ with excitation temperatures 
from $\sim60$ to $\sim600{\rm\:K}$.
Here as representatives, $K=4$ and $K=8$ lines are 
shown as they are less contaminated from neighboring lines.
The emission of sulfur dioxide SO$_2$, another typical hot-core molecule,
with $E_u/k=403\:\K$, also traces the circumbinary disk and the outflow 
cavity (panel c).
However, peaks of these lower-energy transitions of 
CH$_3$CN and SO$_2$ with $E_u/k\la1000\:\K$ 
do not coincide with the positions of sources A and B,
due to self-absorption 
and/or absorption against the compact continuum 
sources in slightly redshifted velocities, indicating that they trace the outer cooler infalling material.
This wide distribution makes it difficult to study the innermost regions of a few hundred au by these lines.

On the other hand, the vibrationally excited transitions
of SO$_2$, CS, and H$_2$O with upper-state energies of
$E_u/k\ga1000\:\K$ trace the innermost region of the 
circumbinary disk and the individual protostellar disks (panels d--f).
In particular, the H$_2$O $v_2=1$ emission with 
$E_u/k=3464\:\K$ is concentrated at the positions 
of sources A and B (panel f; 
the extended emission comes from contamination of 
other lines; see Figure \ref{fig:outer-disk}(d)).
Such a high upper-state energy reflects the high temperature of
protostellar disks in massive star formation at several hundred au.
With lower $E_u$, the SO$_2$ $v_2=1$ and CS $v=1$ lines also trace the
rotation of the circumbinary disk on the 1000 au scale (see below).
We note that \citet{zap19} first reported the CS $v=1$ emission tracing the rotating circumbinary disk,
but its connection to the individual circumstellar disks were not known.

Furthermore, we found that the emissions of NaCl, SiO, and SiS are also concentrated
in the vicinity of the protostars
(panels g--i; again the extended emissions come from contamination of other lines).
These lines are not detected in the 1000 au scale,
although they have the low upper-state energies of $E_u/k<100\K$,
and their critical densities of $\sim10^6$--$10^7{\rm\:cm^{-3}}$ are not high.
This fact indicates that these refractory
molecules are enhanced only in the innermost regions of several hundred au.
It is worth noting that this is the second reported detection of the alkali metal halide, NaCl,
in protostellar systems after the Orion Source I disk \citep{gin19,wri20}.

Appendix Table \ref{tab:line} summarizes the emission lines presented in this Letter.
We note that some other transitions of NaCl, SiO, SiS, and vibrationally excited H$_2$O are also detected,
which will be reported in a future paper.

\begin{figure*}
\begin{center}
\includegraphics[width=\textwidth]{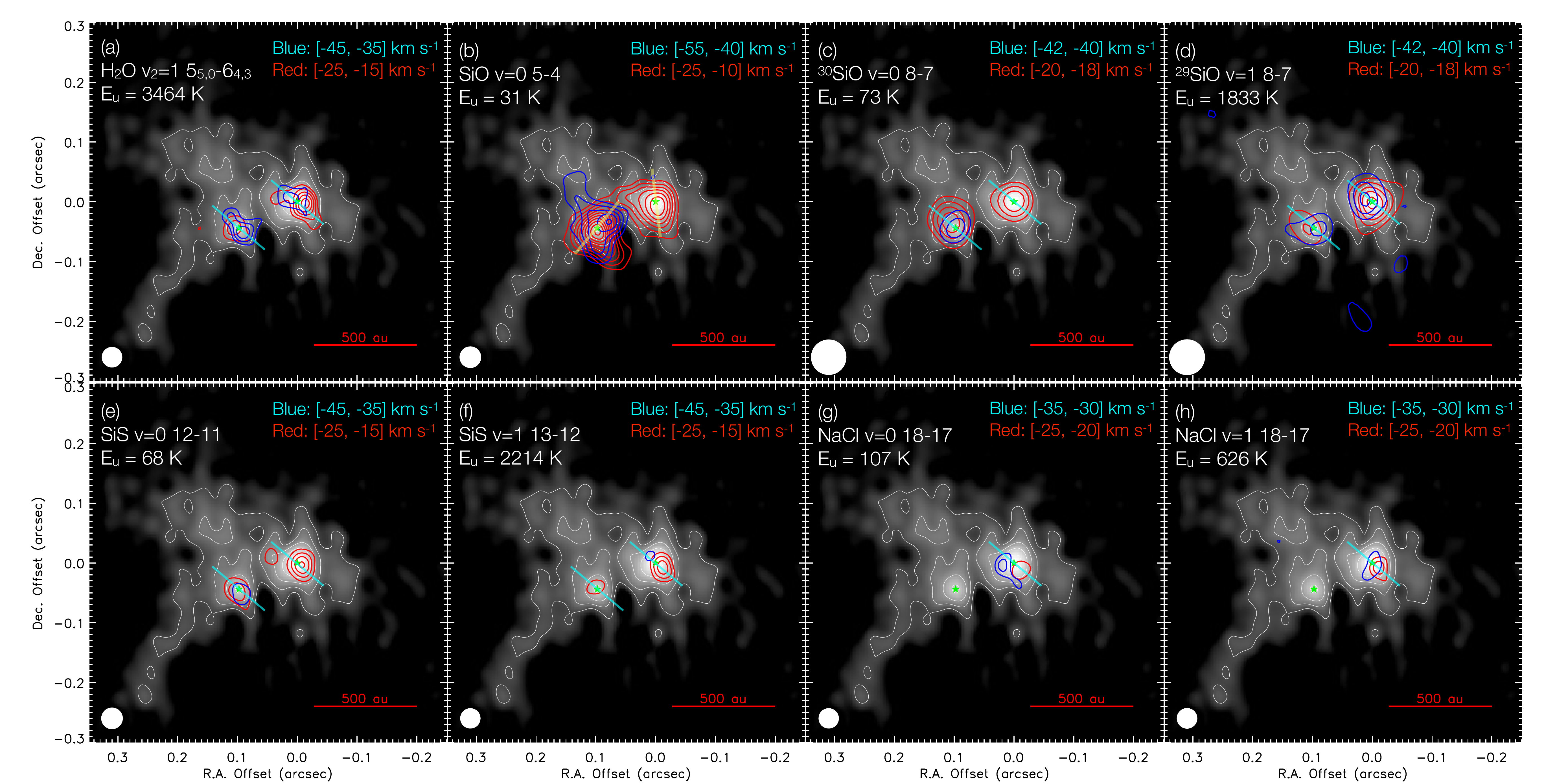}\\
\caption{
Integrated blueshifted and redshifted emission maps of 
selected inner-disk tracing molecular lines (blue and red contours)
overlaid on the 1.3 mm continuum emission (grayscale and white contours).
The molecule names, transitions, upper-state energies $E_u$, and the integrated $\vlsr$ ranges are labeled in each panel.
The stars mark the continuum peaks of sources A and B.
In panels a, b, and d--h, the blue and red contours have lowest contour 
levels of $5\sigma$. In panel c, the blue and red contours have 
lowest contour levels of $15\sigma$ and $10\sigma$, respectively.
The blue and red contour intervals are $2.5\sigma$ in all panels. 
Here $1\sigma=92, 75, 13, 12, 61, 124, 112, 149~\kkms$ in panels a--h, respectively.
The cyan lines are to guide the eye of the tentative orientations of the disk rotation (panels a, and c--e),
and the yellow lines for the outflows (panel b).
}
\label{fig:disk}
\end{center}
\end{figure*}

Using these lines, we can illustrate the kinematics from the 
circumbinary disk to the individual circumstellar disks.
Figure \ref{fig:outer-disk}a presents the moment 1 map of the 
SO$_2$ $v_2=1$ line, showing the rotation of the circumbinary disk 
as reported by \citet{zap15,zap19}.
The systemic velocity of IRAS 16547 is about 
$-31{\rm\:km\:s^{-1}}$ \citep{gar03}.
The rotation direction is consistent with the elongation of the circumbinary structure.
Following \citet{zap15,zap19}, we plot the position-velocity (PV) 
diagrams along the major axis of the circumbinary disk 
(P.A.=$50\degr$), passing between sources A and B (panel b).
The PV diagram of the SO$_2$ $v_2=1$ line shows 
a rotational profile with velocity increasing toward the center.
However, inside $0.1\arcsec$, or $300{\rm\:au}$,
the SO$_2$ $v_2=1$ emission does not show
the high-velocity component, which is expected for the Keplerian disk.
Instead, the SiO emission nicely traces the central high-velocity 
components up to $\Delta \vlsr \simeq\pm30\:\kms$.

Figure \ref{fig:outer-disk}c shows the PV diagrams of the SO$_2$ $v_2=1$ 
and SiO emissions along a slit passing through sources A and B 
(P.A.$=-65\degr$).
This PV diagram is clearly not a simple Keplerian profile inside $0.1\arcsec$,
suggesting that the two protostars dominate the dynamics at this scale.
Not only the SiO line but also the SO$_2$ $v_2=1$ line shows
the high-velocity components (especially in redshifted velocities), associated with sources A and B.
The same is also seen in the PV diagrams of the H$_2$O $v_2=1$ 
emission with P.A.$=-65\degr$ (panel d),
where the hot-water emission prominently shows two circumstellar components.
These indicate that the rotation around the binary system
is smoothly connected to the rotation around the individual protostars.
The two circumstellar components are not quite parallel in the PV diagrams,
judging from the different direction of velocity gradient.
This suggests misalignment of the rotation directions between the twin disks (see below).
We note that the contaminations by other lines are seen,
extending to the southeast direction at $\vlsr\simeq-20$ and $-27\:\kms$.
We have not been able to identify these contamination lines.
This source is rich in complex organic molecules,
and a simultaneous check through the whole spectrum ranges is required for accurate line identification,
which is left for future work.

We note that the blueshifted emission of SiO in source A is missing,
probably due to self-absorption 
(see Figure \ref{fig:spectrum}),
indicating that the SiO emission traces the outflowing material.
However, as opposed to the commonly seen extended SiO emissions
tracing shocked regions along the outflow,
here the compact morphology of SiO and its close
association with the two protostars suggest that
it traces the material just launched from the disks
or the surface layers of the disks,
which can show both rotation and outflowing motions
\citep[e.g., ][]{hir17,mau19,zha19a}.

\begin{figure}
\epsscale{1.2}
\plotone{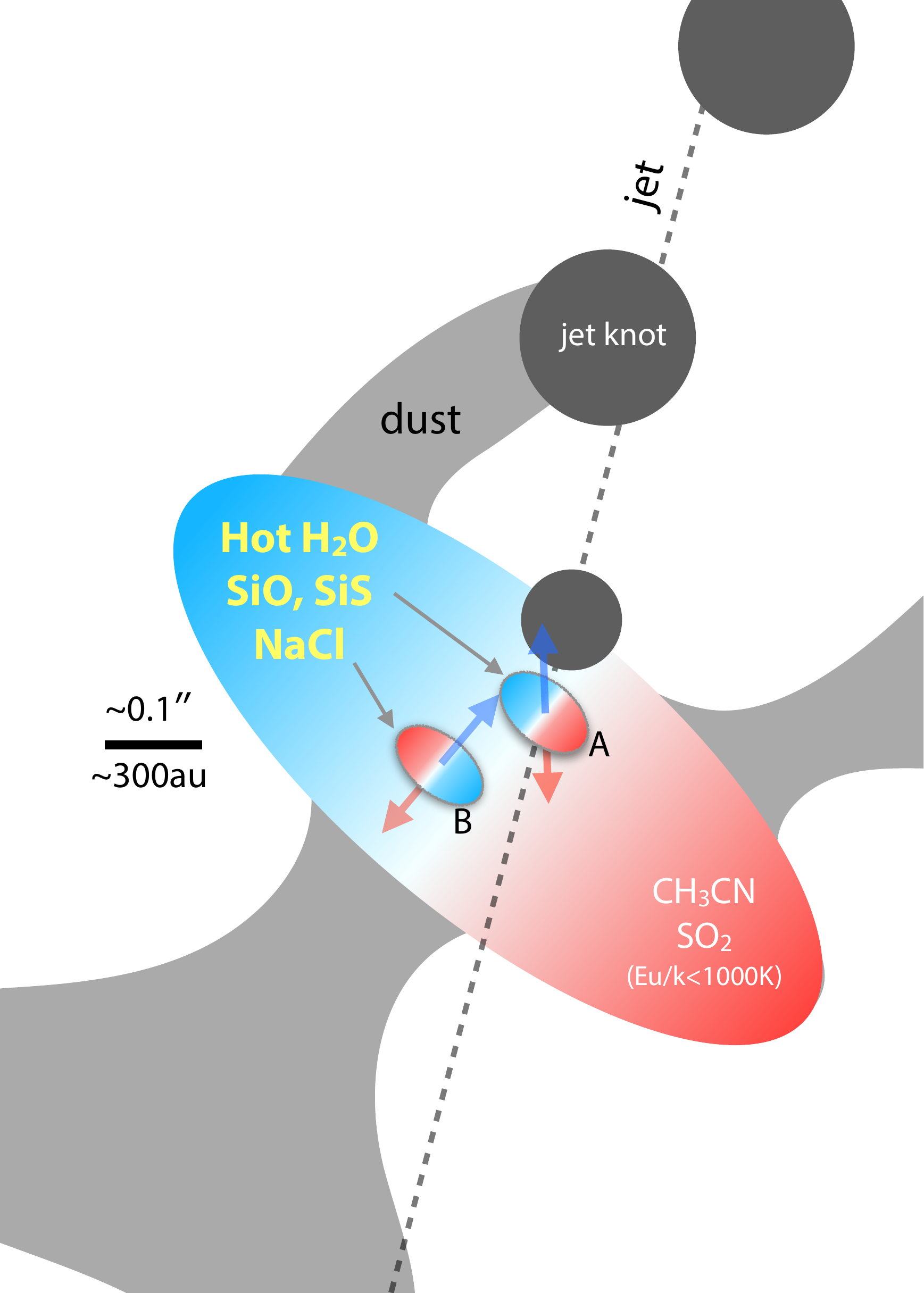}
\caption{
Schematic view of the massive proto-binary in IRAS 16547--4247.
The central twin disks are revealed by high-energy transition H$_2$O lines with $E_u/k>3000{\rm\:K}$,
as well as NaCl and silicon-compound lines that are produced by the destruction of dust grains.
The circumbinary disk, the dusty outflow cavity, and the jet knots are also seen by the new ALMA observations.
The blue and red colors indicate the blueshifted and redshifted emissions from rotation.
The circumstellar disk B is found to be counter-rotating against the disk A and the circumbinary disk.
The outflowing materials from sources A and B are also traced by the SiO emission (blue and red arrows).
\label{fig:schematics}}
\end{figure}

Figure \ref{fig:disk} shows the blue- and redshifted emissions of selected lines of water (panel a), 
silicon-compound (panels b--f), and sodium chloride (panels g--h).
To better resolve the kinematics of the individual circumstellar disks,
for the lines in Band 6, we further improve the resolution to $\sim0.035\arcsec$
($\sim100{\rm\:au}$), by emphasizing data with longer baselines
using a robust parameter of $-0.5$.
For source A, the H$_2$O and NaCl emissions show velocity gradients in a northeast--southwest direction,
which is similar to the rotation direction of the circumbinary disk (P.A.$=50\degr$).
Therefore, we interpret this velocity gradient as the disk rotation of source A.
This orientation is also consistent with the rotating structure traced by water maser
\citep{fra09} around source A.
We note that this disk feature is not affected by the extended contaminations seen
at $\vlsr\simeq-20$ and $-27\:\kms$ in Figure \ref{fig:outer-disk}d,
because they are significantly weakened by emphasizing the longer baselines.
The disk A rotation is more difficult to identify in silicon-compounds emissions because 
these emissions could be blended with the outflowing motion.
The velocity gradients of the $v=0$ emissions of SiO and $^{30}$SiO are ambiguous
due to the strong absorption in the blueshifted component.
For the SiO emission,
comparing the redshifted emission position and the continuum peak position
gives a velocity gradient direction in ${\rm P.A.}\sim10\degr$, 
which could be the outflow direction (or the direction between the outflow and the disk rotation).
If this is the case, resembling the typical star-formation picture,
the outflow direction would be nearly perpendicular to the disk rotation,
and close to parallel with the jet-knot orientation.
On the other hand, the velocity gradient of the vibrationally excited 
$^{29}$SiO line is consistent with the disk A rotation,
as this line is optically thinner than the SiO $v=0$ line due to its rarity and high-excitation state.
For SiS, redshifted components of $v=0$ and $1$ lines roughly follow
the same velocity gradients seen by H$_2$O and NaCl,
suggesting the existence of SiS in the disk.
However, the blueshifted component is missing in the low excitation ($v=0$) map,
probably due to the similar reason for SiO.
The NaCl lines trace the disk components even for the lower excitations,
suggesting NaCl does not exist in the outflow unlike silicon compounds.

In source B, the velocity gradients seen in the emissions of
H$_2$O, $^{30}$SiO, $^{29}$SiO, and SiS $v=0$ are close to parallel to the disk A rotation,
but in the {\it opposite} direction,
suggesting that
the circumstellar disk of source B is rotating in the opposite direction to the disk A and the circumbinary disk.
The high-velocity component of the SiO $v=0$ emission again
shows a gradient perpendicular to the disk rotation,
which may also trace the outflowing motion, similar to source A.

\section{Discussions}\label{sec:discussion}

\subsection{Salt, Silicon Compounds, and Hot Water\\as Disk Probes}\label{sec:tracer}

Based on the new high-resolution ALMA observations,
we identify two groups of molecular lines tracing the innermost 100 au 
scale of the massive binary system IRAS 16547.
The first group is the vibrationally excited ``hot" lines with 
$E_u/k>1000\:\K$.
Especially, the H$_2$O line with $E_u/k>3000\:\K$ nicely traces the 
individual circumstellar disks.
The second group is the refractory molecules, i.e.,
alkali halides (NaCl) and silicon compounds (SiO and SiS) in the case of IRAS 16547.
The lines of refractory species do not necessarily have high excitation of $>1000\:\K$,
but they trace only the innermost regions around the circumstellar disks.
This fact indicates they are released to the gas phase within the disks on the 100~au scale.
The production pathway of these gas-phase refractory molecules
is probably through the destruction of dust grains,
e.g., sputtering under strong shocks or radiation, and thermal desorption from grain surfaces.
SiO and SiS are likely released to the gas phase in the disk and then launched to the outflow,
as they are associated with both the disks and the outflows.
Further multi-line high-resolution observations and their excitation analysis will be
crucial to investigate chemical origin of those species.

Some previous observations have reported similar molecular lines 
tracing the innermost regions of massive protostellar sources.
Orion Source I has been intensively studied as it is the closest 
massive protostar candidate at a distance of $415{\rm\:pc}$.
All the disk-tracing molecules reported here, i.e., H$_2$O, SiO, SiS, and NaCl, 
are detected in Orion Source I's disk and wind launched from the disk \citep{hir12,hir14,hir17,gin19,wri20}.
Moreover, emission lines of aluminum monoxide (AlO),
one of the most refractory materials, are also found to be associated with the central region of 
Orion Source I \citep{tac19}. 
However, Orion Source I might not be the prototypical massive protostar
because it has some peculiar features,
e.g., the lack of an envelope, and the association of extremely strong SiO masers.
Some claimed that Orion Source I could be an evolved star rather than a protostar \citep[see][]{bae18}.
In either case,
the single example could not establish these molecules
as common disk probes of massive star formation.

Recently, in the B-type protostar G339.88--1.26,	
\citet{zha19a} found that the disk and envelope can be disentangled	
not only by kinematics but also by chemical signatures. In particular,	
they found that the Keplerian disk is traced by SiO emission.	
\citet{mau18,mau19} also presented the SiO and vibrationally excited H$_2$O	
emissions tracing a Keplerian disk around the O-type protostar G17.64+0.16	
at $\la100{\rm\:au}$.	
Moreover, the lines of sodium chloride are detected in the disk of G17.64+0.16 (L. Maud 2020, private communication).	
In addition, in both G339.88--1.26 and G17.64+0.16,	
these authors noted that the complex organic molecules and typical hot-core lines (such as CH$_3$CN)	
trace the envelopes rather than the disks,	
similar to our case of IRAS 16547.
Consistent with these previous findings, in the case of IRAS 16547,
the individual circumstellar disks on $100{\rm\:au}$ scale are traced by the 
NaCl, SiS, and vibrationally excited H$_2$O lines.
These studies suggest that hot water, silicon compounds, and alkali halides
could be commonly present in dynamical and hot massive protostellar sources,
and can be used to trace the inner disk and/or the material
just launched from the disk.
Further systematic observations are needed to confirm the common presence
of those molecular lines in massive protostellar sources.
Developing from the conventional hot-core chemistry,
the ``hot-disk" chemistry would be an essential avenue 
for future research of massive star formation.
This work has demonstrated the usefulness of the hot-disk lines for understanding 
the dynamics down to $\sim100{\rm\:au}$ from the massive protostars.
We also note that the lower-energy transitions of refractory molecules are excellent targets for
 future radio observations
by the Square Kilometre Array (SKA) and the Next Generation Very Large Array (ngVLA),
which will be able to resolve the sublimation fronts of solid materials at a 10 au scale.

An additional importance of the hot-disk chemistry around protostars is its unique link to meteoritics.
The oldest materials contained in primitive meteorites, i.e., Ca-Al-rich inclusions (CAIs) and chondrules,
have been sublimated or molten once in the proto-solar disk.
This fact suggests that at least some materials in the pre-solar nebula must be heated to $\ga1500\:\K$,
although protoplanetary disks are typically as cool as few hundred Kelvin in planet-forming 
regions of several au scale \citep[e.g.,][]{bel00}.
Therefore,
how and where CAIs and chondrules formed is still a matter of debate.
Further observations of hot-disk chemistry
could provide important constraints on the gas-phase conditions of refractory species,
and might give unique insights into the formation of high-temperature meteoritic components.

\subsection{The Massive Proto-binary IRAS 16547} \label{sec:iras16547}

Finally, we discuss the unveiled picture of the massive proto-binary IRAS 16547,
and a possible scenario of its origin (see the schematics in Figure \ref{fig:schematics}).
The orbital dynamics could be constrained based on the systemic-velocity difference between two sources \citep{zha19b}.
We find the velocity difference is as small as $\Delta v_{\rm lsr}\la2{\rm\:km\:s^{-1}}$ based on the available inner-disk tracing lines (Figure \ref{fig:spectrum}).
If the two protostars are coplanar with the circumbinary disk,
and orbiting the same circular path following the Keplerian profile of the circumbinary disk
\citep[the enclosed mass of $25\:M_\odot$; the inclination of $55\degr$ by][]{zap19},
the expected velocity difference is about $4{\rm\:km\:s^{-1}}$.
The fact that the observed $\Delta v_{\rm lsr}$ is smaller than the 
simple Keplerian velocity suggests that the protostars are gravitationally bound.

The ionized state of surrounding environments provide hints of the 
evolutionary stage of massive protostars \citep{KT16,ros19,zha19c}.
Based on the free-free fluxes at 3 mm,
we estimate ionizing photon rates of
$9.6\times10^{45}{\rm\:s^{-1}}$ and $4.3\times10^{45}{\rm\:s^{-1}}$ for 
sources A and B, respectively (Appendix \ref{sec:appB}).
Note that those are upper limits as we ignore the contribution 
from dust emission.
The estimated ionizing-photon rates are several orders of magnitude lower
than that of zero-age main-sequence (ZAMS) stars with $>2\times10^4\:L_\odot$ \citep{dav11}, 
suggesting that the binary stars are at protostar phase with large radii of $\sim20\:R_\odot$.
The evolutionary calculations of protostars proposed that
such large radii of massive protostars are the consequence of high accretion rates of $\ga10^{-4} M_\odot{\rm\:yr}^{-1}$
\citep[e.g.,][]{hos09,hae16}.

Although the free-free emission has been detected in IRAS 16547 at radio wavelengths \citep{rod05,rod08},
we do not detect hydrogen recombination lines such as H40$\alpha$ and H42$\alpha$.
The non-detection of recombination lines can be explained by the line broadening with a width of
$\ga100{\rm\:km\:s^{-1}}$ (Appendix \ref{sec:appB}),
which is consistent with the presence of jets seen in the 3 mm continuum.
Additionally, the northern jet knots are located with approximately equal intervals of $0.4\arcsec$, 
or $1000{\rm\:au}$,
which may be evidence of periodic accretion induced by a hidden companion or 
some disk instability around source A.
Assuming a total mass of $20~M_\odot$ and a jet velocity of $100~\kms$ on the sky plane,
we estimate this hidden companion should have a period of $50~\yr$ and a semi-major axis of $40~\au$.
The proper motion of the jet knots would be detectable by follow-up observations with similar resolutions, 
which will provide important clues for testing the companion's presence and 
for understanding jet launching and precession \citep{rod08}.

The two protostars have similar continuum fluxes and line emissions, 
and look coplanar with the circumbinary disk.
Those features superficially link to the disk fragmentation as the origin of the binary system \citep[e.g.,][]{kru09}.
A puzzling finding, however, is the tentative detection of the counter-rotating disks (Figure \ref{fig:disk}),
which are difficult to form by disk fragmentation.
An alternative mechanism is turbulent fragmentation at the molecular cloud-core scale 
\citep[e.g.,][]{off10,bat12,kuf19}.
Although their birthplaces may be distant,
some pairs of protostars can migrate to as close as $\la100{\rm\:au}$, forming binary systems.
The presence of turbulence leads to the random rotation of protostellar disks, 
which remains even after the migration \citep{off16}.
The turbulent fragmentation scenario would go well with
the small-cluster nature of IRAS 16547 on the scale of $\la0.1{\rm\:pc}$,
seen in the misalignments of several outflows and jets \citep{hig15}.
However, considering the actual origin of binary systems could be much more complicated,
e.g., the combination of both fragmentation processes \citep{rosen20}
and the dynamical interactions with highly eccentric orbits \citep{sai20},
it is difficult to conclude the formation process based on the currently available information.
We want to emphasize that the detection of the counter-rotation is still tentative,
and follow-up high-resolution observations are required to conclude the disk orientations of IRAS 16547.

\section{Summary}\label{sec:summary}
We report the dynamical and chemical structures of the massive 
proto-binary system IRAS 16547--4247
using $0.05\arcsec$-resolution ALMA observations at 3, 1.3, and 
0.85 mm.
We propose that (1) the lines of destructed-dust molecules, such as 
alkali metal halides (e.g., NaCl) and silicon compounds (e.g., SiO and SiS), and
(2) the high-excitation water line with $E_u/k>3000{\rm\:K}$ 
can act as good tracers for
investigating dynamics of the innermost region of massive star formation
at a scale of $\la100{\rm\:au}$.

Figure \ref{fig:schematics} shows the schematic view of the proto-binary
IRAS 16547 presented in this study.
In the scale of $1000{\rm\:au}$,
the rotation of the circumbinary disk
is revealed by emission lines of typical hot-core molecules,
such as CH$_3$CN and SO$_2$, with upper-state energies of 
$E_u/k\simeq100$--$1000{\rm\:K}$ \citep[][]{zap19}.
However, these lines cannot trace well the protostellar disks
at a 100 au scale.
Instead, we found that some molecular lines, including vibrationally excited water, silicon compounds,
and sodium chloride, exclusively trace the individual circumstellar disks.
The detection of vibrationally excited lines in H$_2$O, SiO, SiS, and NaCl
with upper-state energy as high as $>2000$--$3000{\rm\:K}$ indicates
a very high temperature in the innermost disks.
Because sodium chloride and silicon compounds are produced through the destruction 
of dust grains in the dynamical disks,
their emissions are seen only in the vicinity of protostars,
even for the lower-energy transitions with $E_u/k<100{\rm\:K}$.
Using these new disk probes, we analyzed the disk kinematics
and tentatively discovered that the twin disks are counter-rotating.
The pair of the counter-rotating disks might suggest that
the binary system is formed via turbulent fragmentation at the cloud-core scale
rather than disk fragmentation.
However, more observations are needed to confirm the rotation directions of disks.

Notably, this is the second reported detection of salt in protostellar systems
after the case of the disk of Orion Source I \citep{gin19},
and also one of few massive protostellar disks associated
with high-energy transition water and silicon compounds \citep[e.g.,][]{mau18,mau19,zha19a}.
These new results suggest these ``hot-disk" lines
may be common in innermost disks around massive protostars
and can be detected in high-resolution observations.
Such ``hot-disk" chemistry has great potential
for the future research of massive star formation.

\acknowledgments

The authors thank T. Nakamoto, M. Kobayashi,	
L. Maud, and A. Ginsburg for fruitful discussions.
The authors also thank the anonymous referee
for providing insightful comments.
This Letter makes use of the following ALMA data:
ADS/JAO.ALMA\#2016.1.00992.S.,
and ADS/JAO.ALMA\#2018.1.01656.S.
ALMA is a partnership of ESO (representing its member states),
NSF (USA) and NINS (Japan), together with NRC (Canada),
MOST and ASIAA (Taiwan), and KASI (Republic of Korea),
in cooperation with the Republic of Chile.
The Joint ALMA Observatory is operated by ESO, AUI/NRAO and NAOJ.
This research is supported by NAOJ ALMA Scientific Research grant No. 2017-05A (K.E.I.T. and K.T.),
ERC project MSTAR, VR grant 2017-04522 (J.C.T),
RIKEN Special Postdoctoral Researcher Program (Y.Z.),
JSPS KAKENHI grant Nos. JP19H05080, 19K14760 (K.E.I.T.),
19K14774 (Y.Z.), 17K05398 (T.H.),
19H05082, 19H01937 (K.M.),
16H05998, 17KK0091, 18H05440 (K.T.)
and 20K14533 (S.O.).



\software{
CASA \citep[\url{http://casa.nrao.edu},][]{McMullin07}, 
the IDL Astronomy User's Library \citep[\url{https://idlastro.gsfc.nasa.gov,}][]{lan95}.}

\clearpage
\appendix

\restartappendixnumbering

\setcounter{table}{0}
\setcounter{figure}{0}
\section{Information of the Observations} \label{sec:appA}
Table \ref{tab:obs} summarizes the information of the observations
of Band 3, 6 (project: 2018.1.01656.S), and 7 \citep[project 2016.1.00992.S,][]{zap19}.
As listed, similar high resolutions of $\sim0.05\arcsec$ are achieved for all wavelengths.

\begin{table} 
\scriptsize
\begin{center}
\caption{Information of the Observations}\label{tab:obs}
\begin{tabular}{ccccccccc}
\hline
\hline
Band & Obs. Date (Total Time) & \# of Ant. & 
Baseline Range & Phase Cal. & Flux/Bandpass Cal. & 
$\theta_{\rm beam}$ \footnote{The synthesized beams for the continuum images. A robust weighting parameter of 0.5 is used for Band 3 and 6 data, and a robust weighting parameter of -0.5 is used for Band 7 data.} & $\theta_{\rm MRS} \footnote{The maximum recoverable scales.} $ \\
\hline
3 & June 9, 2019 (90 min) & 45 & 83.1~m $-$ 16.2~km &
J1706-4600 & J1617-5848 &  
$0.048\arcsec\times0.046\arcsec$ & $0.73\arcsec$ \\
6 & July 15, 2019 (78 min) & 42 & 138.5~m $-$ 8.5~km &
J1706-4600 & J1427-4206 &  
$0.055\arcsec\times0.038\arcsec$ & $0.75\arcsec$ \\
7 & August 21 $-$22, 2019 (100 min) & 44 & 21.0~m $-$ 3.7~km & 
J1636-4102 & J1617-5848/J1427-4206 & 
$0.056\arcsec\times0.046\arcsec$ & $0.84\arcsec$ \\
\hline
\end{tabular}
\end{center}
\end{table}

\setcounter{table}{0}
\setcounter{figure}{0}
\section{Estimations of Protostellar Properties} \label{sec:appB}

\begin{figure}
\begin{center}
\includegraphics[width=0.8\textwidth]{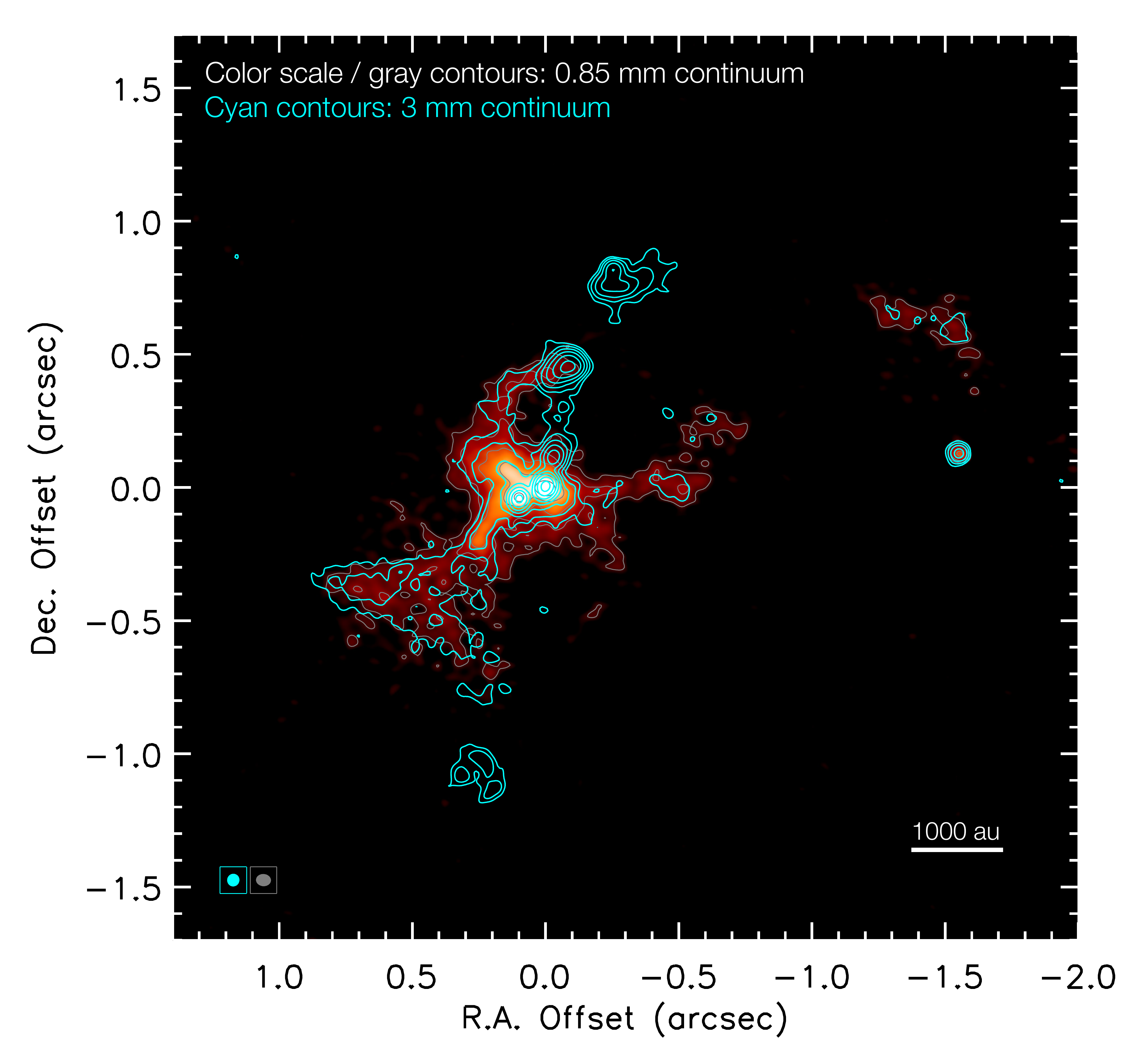}\\
\end{center}
\caption{
Same as Figure \ref{fig:cont},
but showing the 0.85 mm continuum image in color scale and gray contours.
The contour levels are $5\sigma\times2^n$ ($n=0,1,...$), 
with $1\sigma=1.6~\K$ ($0.40~\mJybeam$).
The synthesized beam is $0.056\arcsec\times0.046\arcsec$
($\pa=85.9^\circ$) for the 0.85 mm continuum image (shown at the lower-left corner).
\label{fig:cont2}}
\end{figure}

We estimate the properties of the protostars based on results of multi-wavelength ALMA observations.
The disk masses can be estimated using the dust flux of $S_{\nu,d}$,
\begin{eqnarray}
M_{\rm disk} = \frac{D^2\Omega}{\kappa_{\nu,d}} \log \left( 1- \frac{S_{\nu,d} }{B_\nu(T_d)\Omega} \right)^{-1}, \label{eq:mass} 
\end{eqnarray}
where $D=2.9{\rm\:kpc}$ is the distance to IRAS 16547,
$\Omega$ is the solid angle of the integrated region,
$\kappa_{\nu, d}$ is the dust opacity per gas mass,
and $B_\nu(T_d)$ is the Planck function at the dust temperature $T_d$, respectively.
We utilize the 0.85 mm fluxes of $S_{\nu,d}=84$ and $60{\rm\:mJy}$ within $0.05\arcsec$ for sources A and B,
because the dust emission should dominate at this wavelength (Figure \ref{fig:cont2}).
The dust temperatures in the disks are uncertain from the currently available data,
but the peak brightness temperature of $340\:\K$ at 0.85 mm suggests the high temperature of $T_d\ga350\:\K$.
Here we assume the dust temperature range of $T_d=350$--$500{\rm\:K}$.
Considering the physical condition of the disks,
we apply the dust opacity of $\kappa_{\nu, d}=0.097{\rm\:cm^2\:g^{-1}}$
for the coagulated dust model in the high density of $10^8{\rm\:cm^{-3}}$ without the ice mantle
\citep{oss94} (a gas-to-dust ratio of 100 is assumed).
We evaluate the disks masses as $0.033$--$0.19\:M_\odot$ and $0.019$--$0.035\:M_\odot$ for sources A and B, respectively,
at the dust temperature range of $T_d=350$--$500{\rm\:K}$.

The free-free emissions at radio wavelengths are observed at the center of IRAS16547 \citep{rod05,rod08},
suggesting the existence of photoionized regions.
Under the assumption of the optically thin free-free emission,
we can evaluate the ionizing-photon rates of the protostars \citep{sch16},
\begin{eqnarray}
S_{\rm ion}=4.771\times10^{42} \left(\frac{S_{\nu,{\rm ff}}}{\rm Jy}\right) \left(\frac{T_e}{{\rm K}}\right)^{-0.45}
\left(\frac{\nu}{\rm GHz}\right)^{0.1} \left(\frac{D}{\rm pc}\right)^{2}
{\rm\:s^{-1}},
\end{eqnarray}
where $T_e$ is the electron temperature, which we use the typical values of $8000{\rm\:K}$ \citep{ket08}.
We adopt the 3 mm continuum for the free-free fluxes,
which is the upper limit because the dust emission would contribute.
Based on $S_{\nu,{\rm ff}}=36$ and $17{\rm\:mJy}$ within $0.05\arcsec$,
we evaluate the ionizing-photon rates of $9.7\times10^{45}{\rm\:s^{-1}}$ and $4.3\times10^{45}{\rm\:s^{-1}}$
for sources A and B, respectively.
The evaluated rates are orders of magnitude lower than ZAMS stars with luminosities $>2\times10^4\:L_\odot$ \citep{dav11}
(note the total luminosity of IRAS 16547 is $\sim10^5\:L_\odot$),
confirming that sources A and B are still at the protostellar phase with large radii.
Assuming the bolometric luminosities are $5\times10^{4}\:L_\odot$,
we estimate the stellar radii of $16\:R_\odot$ and $17\:R_\odot$ for sources A and B, respectively \citep{KT16}.
Such large radii of the massive protostars suggest
that both protostars have grown with high accretion rates $\ga10^{-4} M_\odot{\rm\:yr}^{-1}$
\citep[e.g.,][]{hos09,hae16}.

We do not identify the hydrogen recombination lines of H26$\alpha$, H30$\alpha$, H40$\alpha$, and H42$\alpha$
(353.6227, 231.9009, 99.0230, and 85.6884 GHz, respectively),
which suggests a broadening effect due to the high-velocity jets.
Under the assumption of the optically thin and local thermal equilibrium (LTE) conditions of the recombination lines,
we can estimate the the ratio of hydrogen recombination lines to the free-free continuum  \citep{ang18},
\begin{eqnarray}
\frac{I_{\nu,{\rm HRL}}}{I_{\nu, {\rm ff}}}=0.19 \left(\frac{\nu}{\rm GHz}\right)^{1.1} \left(\frac{T_e}{10^4{\rm\:K}}\right)^{-1.1}
\left(\frac{\Delta V}{\rm km\:s^{-1}}\right)^{-1} \left(1+Y^+\right)^{-1},
\end{eqnarray}
where $Y^+$ is the ratio of the He$^+$ and H$^+$ column densities (we use the typical value of $Y^+=0.08$).
Again, we adopt the 3 mm continuum flux as the free-free emission.
Taking into account the peak intensity $I_{\nu,{\rm ff}}=7.2~\mJybeam$
and the rms noise of $0.4~\mJybeam$,
we estimated that the line width would be $\Delta V\ga100~\kms$
for the non-detection with $5\sigma$ level, i.e., $I_{\nu,{\rm HRL}}<2~\mJybeam$.
This wide width is consistent with the presence of the jets.
We note that this is a conservative limit because the 3 mm flux also contains the dust emission.

\setcounter{table}{0}
\setcounter{figure}{0}
\section{Information of the Presented Molecular Lines} \label{sec:appC}

\begin{deluxetable}{lcrrr}
\tablecaption{Molecular Lines Presented in This Letter \label{tab:line}}
\tablewidth{0pt}
\tablehead{
\colhead{Molecule} & \colhead{Transition} & \colhead{Frequency} & \colhead{$E_u/k$} & \colhead{$S\mu^2$}\\
\colhead{} & \colhead{} & \colhead{(GHz)}  & \colhead{(K)} & \colhead{(D$^2$)}
}
\decimals
\startdata
   H$_2$O & $5_{5,0}$--$6_{4,3}$ $(v_2=1)$ & 232.6867000 & 3463.6 & 1.079 \\
            SiO & 5--4 $(v=0)$ & 217.1049800 & 31.3 & 48.00 \\
$^{30}$SiO & 8--7 $(v=0)$ & 338.9300437 & 73.2 & 76.81 \\
$^{29}$SiO & 8--7 $(v=1)$ & 340.6118622 & 1832.6 & 77.77 \\
            SiS & 12--11 $(v=0)$ & 217.8176630 & 68.0 & 12.55 \\
           SiS & 13--12 $(v=1)$ & 234.8129678 & 1150.6 & 169.5 \\
            NaCl & 18--17 $(v=1)$ & 232.5099753 & 626.0 & 1478 \\
            NaCl & 18--17 $(v=0)$ & 234.2519153 & 106.9 & 1458 \\
\hline
   CH$_3$CN & $12_4$--$11_4$  & 220.6792874 & 183.2 & 328.2 \\
   CH$_3$CN & $12_8$--$11_8$  & 220.4758078 & 525.8 & 205.1 \\
        SO$_2$ & $28_{3,25}$--$28_{2,26}$  $(v=0)$& 234.1870566 & 403.2 & 55.09 \\ 
        SO$_2$ & $26_{3,23}$--$26_{2,24}$ $(v_2=1)$& 216.7585584 & 1096.5 & 54.13 \\ 
                CS & 7--6 $(v=1)$& 340.3979569 & 1896.6 & 26.24 \\ 
\enddata
\tablecomments{Line information of H$_2$O is taken from the Jet 
Propulsion Laboratory (JPL) line database \citep{pick98},
the information of other lines are taken from the Cologne Database for Molecular Spectroscopy molecular line catalog (CDMS) \citep{mul05}.
The molecular lines listed above the horizontal line between NaCl and CH$_3$CN
trace well the 100 au scales of IRAS16547
(see Figures \ref{fig:disk} and \ref{fig:spectrum}).
}
\end{deluxetable}

\begin{figure}
\begin{center}
\includegraphics[width=1\textwidth]{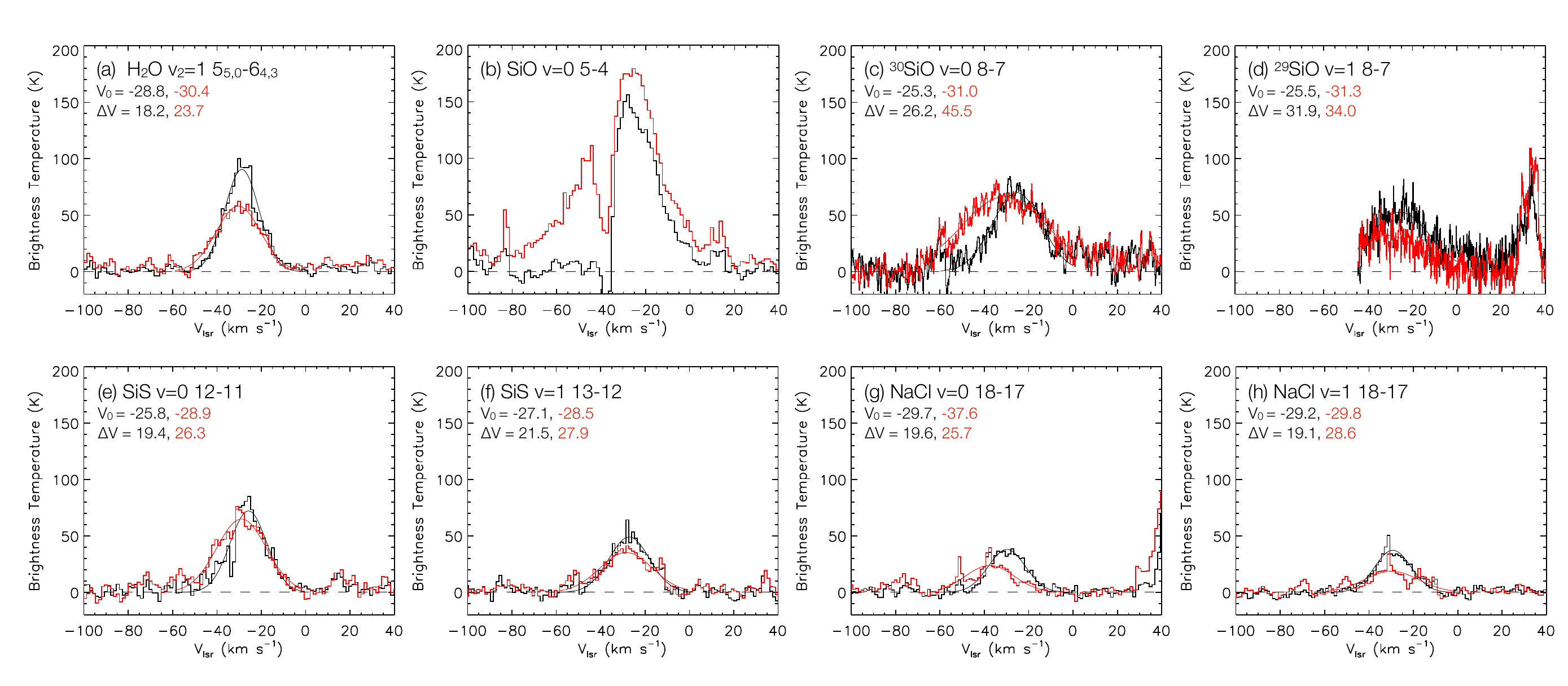}\\
\end{center}
\caption{
Line spectra of water, silicon compounds, and sodium chloride at the continuum peak positions of sources A (red) and B (black).
The fitted Gaussian functions are also displayed with their central and FMHW velocities in the unit of $\mbox{km s}^{-1}$, $V_0$ and $\Delta V$,
except for the SiO (5--4) emission with the strong absorption.
\label{fig:spectrum}}
\end{figure}

Table \ref{tab:line} summarizes the emission lines presented in this study.
We particularly discuss the detection of sodium chloride, silicon compounds, and hot water as the disk probes at the 100 au scale.
Figure \ref{fig:spectrum} shows the spectra of these emissions at the continuum peaks of sources A and B.
The fitted Gaussian functions of each profile are also presented,
except for the SiO (5--4) emission, which has strong absorption.
The systemic velocity of IRAS 16547 is about $-31{\rm\:km\:s^{-1}}$ \citep{gar03}.
As seen in Figure \ref{fig:disk},
the individual disks are traced particularly well
by the emissions of H$_2$O, NaCl $(v=0,~1)$, and SiS $(v=1)$ for source A,
and H$_2$O, and SiS $(v=0)$ for source B.
These lines have quasi-Gaussian shapes with the FMHWs of $18$--$26{\rm km~s}^{-1}$.
We do not identify a clear line-of-sight velocity difference between the two protostars,
i.e., $\Delta v_{\rm lsr}\la2{\rm\:km\:s^{-1}}$,
which indicates that the binary system is gravitationally bound
(Section \ref{sec:iras16547}).
The water line is contaminated by the other lines
at around $\vlsr=-20$ and $-27~\kms$ (see also Figure \ref{fig:outer-disk}d),
which we have not been able to identify.
On the other hand,
the SiO (5--4) emission traces not only the disks but also the outflows.
Thus, it is broader with clear absorption features in the blueshifted side
(see also Figure \ref{fig:mom0}h and i).
Similar features are also seen in some of the other silicon-compound lines,
e.g., SiS $(v=0)$ and $^{30}$SiO of source A.
The deeper absorptions in the silicon-compound lines in source A
suggests its stronger outflow than the outflow B.



\end{document}